\documentclass[5p,number, sort&compress, final]{elsarticle}
\usepackage{amsmath}
\usepackage{amssymb}
\usepackage[separate-uncertainty=true,
 range-units = single,
 range-phrase = -
]{siunitx}

\newcommand{\EeV}{\exa\electronvolt}
\DeclareSIUnit\parsec{pc}
\DeclareSIUnit\lightyear{ly}
\DeclareSIUnit\gauss{G}
\DeclareSIUnit\Sigma{$\sigma$}
\DeclareSIUnit\year{yr}
\DeclareSIUnit\years{yr}
\DeclareSIUnit\erg{erg}

\usepackage{soul}
\usepackage[backgroundcolor=yellow, linecolor=black]{todonotes}

\usepackage{graphicx}
\begin{document}

\title{Energy Spectrum of Fast Second Order Fermi Accelerators as Sources of Ultra-High-Energy
Cosmic Rays}

\author[VUB]{Tobias Winchen\corref{cor1}}
\cortext[cor1]{Corresponding author}
\address[VUB]{Vrije Universiteit Brussel, Pleinlaan 2, 1050 Brussels, Belgium}

\ead{tobias.winchen@rwth-aachen.de}
\author[VUB]{Stijn Buitink}

\date{\today}

\begin{abstract}
	Stochastic acceleration of cosmic rays in second order Fermi processes is
	usually considered too slow to reach ultra-high energies, except in specific
	cases. In this paper we present the energy spectrum obtained from second
	order Fermi acceleration in highly turbulent magnetic fields as e.g.\ found in
	the outskirts of AGN jets in situations where it can be sufficiently fast to
	accelerate particles to the highest observed energies. We parametrize the
	resulting non-power-law spectra and show that these can describe the cosmic
	ray energy spectrum and mass-composition data at the highest energies if
	propagation effects are taken into account.
\end{abstract}


\maketitle

\section{Introduction}
Cosmic rays are observed with energies from
approximately~\SI{10}{\giga\electronvolt} up to energies above~\SI{100}{\EeV}
with a distribution commonly parametrized as a power-law. The spectral index of
the power-law changes at a few certain energies, in particular the `knee' at
approximately $\SI{1d15}{\electronvolt}$ and the `ankle' at approx.\
$\SI{5d18}{\electronvolt}$. Beyond approx.\ $\SI{4d19}{\electronvolt}$ the flux
is strongly suppressed compared to a simple power-law. The definite origins of
cosmic rays and the spectral features knee, ankle, and cut-off are still
unknown, but in the prevailing models cosmic rays with the highest energies
are of extragalactic origin while cosmic rays with the lowest energies are
accelerated in sources within the galaxy~(e.g.~\cite{Kampert2014}).

It is typically assumed that the energy spectrum of the cosmic rays at their
sources has also to be a single power-law $dN /dE \propto E^{-\gamma}$, maybe
with some cut-off at high energies, in order to obtain the power-law observed
at Earth~(e.g.~\cite{Kotera2011}).  A power-law is the natural result of a
stochastic acceleration of particles, when in every acceleration cycle a
constant fraction of particles is lost. Notably, acceleration in relativistic
shocks via the so called first order Fermi mechanism is considered as an
acceleration mechanism that generates a power-law.  In the shock acceleration
model, a particle traverses a shock front multiple times and is efficiently
accelerated, as it gains energy in every cycle.  However, this mechanism has
some difficulties, as the particles have to return to the shock many times
which require special conditions in the shock environment, a high injection
energy of the particles into the accelerating region, or multiple
shocks~\cite{Colgate1994, Blasi2012}.

From first order Fermi acceleration in diffusive shocks, a soft spectral index
of the injection power-law of $\gamma \geq 2$ is expected~\cite{Marcowith2016}.
However, it has been recently recognized by the Pierre Auger
Collaboration~\cite{Matteo2015, Aab2016l} and others~\cite{Allard2008, Aloisio2014,
Unger2015}, that the spectrum and composition data at the highest energies can
not be well described by sources with a power-law spectrum and a soft spectral
index; instead, hard spectra with $\gamma < 1.5$, or even $\gamma < 0$
depending on the model for the infrared background,  are required to fit the
data with a single power-law.

These problems do not arise in the original `second order' acceleration
mechanism proposed by Fermi~\cite{Fermi1949}, where particles gain energy in
collisions with magnetic clouds.  Here energy losses are  frequent due to
tail-on collisions, the average energy gain per collision is only $\Delta E / E
= \frac{4}{3} \beta^2$.  For velocities of the scatter centers $\beta \ll 1$
this second order acceleration thus requires a large number of scatter events
to significantly gain energy.  With the originally considered low rate of
scatters with slowly moving magnetic clouds at Galactic distances, the
mechanism is thus not fast enough to reach the highest observed energies within
the lifetime of the respective sources.

To reach the highest energies via second order Fermi acceleration nevertheless,
a high velocity of the scatter centers and/or a low mean free path length
between scatter events is required.  In the relativistic limit $\beta \approx
1$ the second order mechanism becomes as efficient as the first order
mechanism. This scenario has been discussed in the
literature~\cite{Pelletier1999} in particular in the context of acceleration
inside GRBs. In general it also yields a power-law emission spectrum.

Here we focus on the second case of scattering with a short mean free path.
Such conditions may be found in turbulent magnetic fields; for example,
scattering on Alfv\'en waves in the radio lobes of AGNs, and in particular
Centaurus A,  has been considered~\cite{OSullivan2009, Fraschetti2008,
Hardcastle2009}.  In this paper we will first discuss our simulations setup and
the spectrum we obtain with it from second order Fermi accelerators. We will then use a
parametrization of the acceleration spectrum to fit observational data with the
obtained acceleration spectrum taking propagation effects into account. The
parameters obtained in the fit, the assumptions of the simulation as well as
potential acceleration environments are then discussed before we draw our final
conclusions.

\section{Cosmic Ray Diffusion as Random Walk}
The diffusive propagation of particles that scatter at irregularities
$\frac{B}{\delta B}$ of magnetic fields with strength $B$, can be described as
a random walk with step length $\lambda \propto D$ for diffusion coefficient
$D$.  The value of the diffusion coefficient $D$, its dependence of the energy
of the particle, and the underlying principles of magnetic turbulences in
astrophysical plasmas are a complex subject of ongoing intensive
studies~\cite[e.g.][]{Lazarian2012}.

A simplistic approach to diffusion of charged particles in magnetic fields is quasi-linear
theory, where only
slow evolution of the plasma is assumed~\cite{Schlickeiser1989}.  Here the mean free path $\lambda$ of a
particle with energy $E$ and charge $Z$ in a field with turbulence spectrum
$\frac{k}{k_{\min}}^{-q}$ is
\begin{equation}
	\lambda = {\left(\frac{B}{\delta B}\right)}^2 {\left(R_G\;  k_{\min}\right)}^{1-q} R_G \equiv \lambda_0 {\left( \frac{E}{\SI{1}{\EeV}}\frac{1}{Z} \right)}^{2-q}
	\label{eq:MeanFreePath}
\end{equation}
where $R_G = \frac{E}{B Z}$ is the gyro-radius of the
particles~\cite{Scalo2004}. This approximation for diffusive transport
of particles with gyro-radius not larger than the correlation
length of the field has been successfully used in several studies of
particle acceleration and transport~\cite[e.g.][]{OSullivan2009, Casse2001}.

However, this formalism is too simplified to provide an accurate quantitative
description of propagation in turbulent fields. In particular, turbulence is
anisotropic~\cite{Goldreich1995} and scattering on small length scales is
dominated by fast-modes instead of Alfv\'enic modes~\cite{Yan2002}. For particles
with small gyroradius smaller than the turbulence injection scale of the magnetic field the
propagation is dominated by fast super-diffusion~\cite{Lazarian2014}.

Consequently the numerical value of the calculated diffusion coefficient for a
given set of plasma parameters is not correct and also the dependency of the
scatter frequency on the particle rigidity is not necessarily uniform over all
scales as suggested by QLT.  In particular the scatter frequency increases at
low energies, eventually reaches a plateau or peak, and decreases
again~\cite{Yan2004}. Nevertheless, the increase of the step length can be
parametrised by a power-law as is expected in some
non-linear theories~\cite{Shalchi2009} and seen in
simulations~\cite{Beresnyak2011}, at least above a certain threshold rigidity.
Therefore, we use as simplified model for second order Fermi acceleration a
random walk with a step length given by eq.~\ref{eq:MeanFreePath} in our
simulation  and consider $\lambda_0$ and $q$ free fit parameters. We further
assume that particles are injected  by a pre-accelerator above a threshold energy where these
assumptions are valid.

\section{Simulation}
To simulate particle acceleration by magnetic scattering we  follow a
test-particle approach. Neglecting the back reaction of the particle on the
magnetic field is valid here, because we are concerned only with the small
fraction of cosmic rays at the highest energies. We demonstrate later that the
energy transferred  to UHECR from the field is much smaller than the energy
transferred to radio emission in potential sources, thus justifying this
approach.

We implemented the simulation as a dedicated module within the CRPropa
framework~\cite{Batista2016}.  The module scatters the particles  into a random
direction in the rest-frame of a scattering center moving with velocity $\beta$
in the laboratory frame after propagating a distance $d$ randomly chosen
according to an exponential distribution with mean free-path $\lambda$
according to eq.~\ref{eq:MeanFreePath}.  We assume isotropic movements of the
cosmic rays and scatter centers in the rest frame of the accelerating region.
The distribution of angles $\theta$ between the direction of the particle and
the scatter center is thus $dN / d\theta \propto 1 - \beta \cos{\theta}$. In
every scatter event a random direction is chosen according to this
distribution. As by this head-on collisions are more likely than tail-on
collisions,  the particle gains on average energy as predicted by second order
Fermi acceleration.

The particles are injected with energy $E_{\text{inj}}$ into the center of a
spherical simulation volume with radius $R$, corresponding to a weak
pre-acceleration e.g.\ in the inner parts of the AGN jet. Every  particle
propagates linearly until it is scattered as described above.  We
stop the simulation when the particle leaves the simulation volume or is
decelerated below a lower energy threshold $E_{\text{low}}$.  Injection energy
and lower energy threshold are chosen as $E_{\text{inj}} =
\SI{100}{\tera\electronvolt}$ and $ E_{\text{low}} = \SI{1}{\tera\electronvolt}$
as in CRPropa currently only the highly relativistic case $E \gg m_0 c^2$ can
be calculated. The simulation does not depend on the absolute scale of
$\lambda_0$ and $R$, allowing us to investigated different values of their
ratio $\lambda_0 / R$, $\beta$, and $q$ only. For acceleration of particles to
energies much higher than the injection energy, the result is independent of
the injection energy and thus on the spectrum of the pre-accelerator.
\begin{figure}[tb]
	\includegraphics[width=\columnwidth]{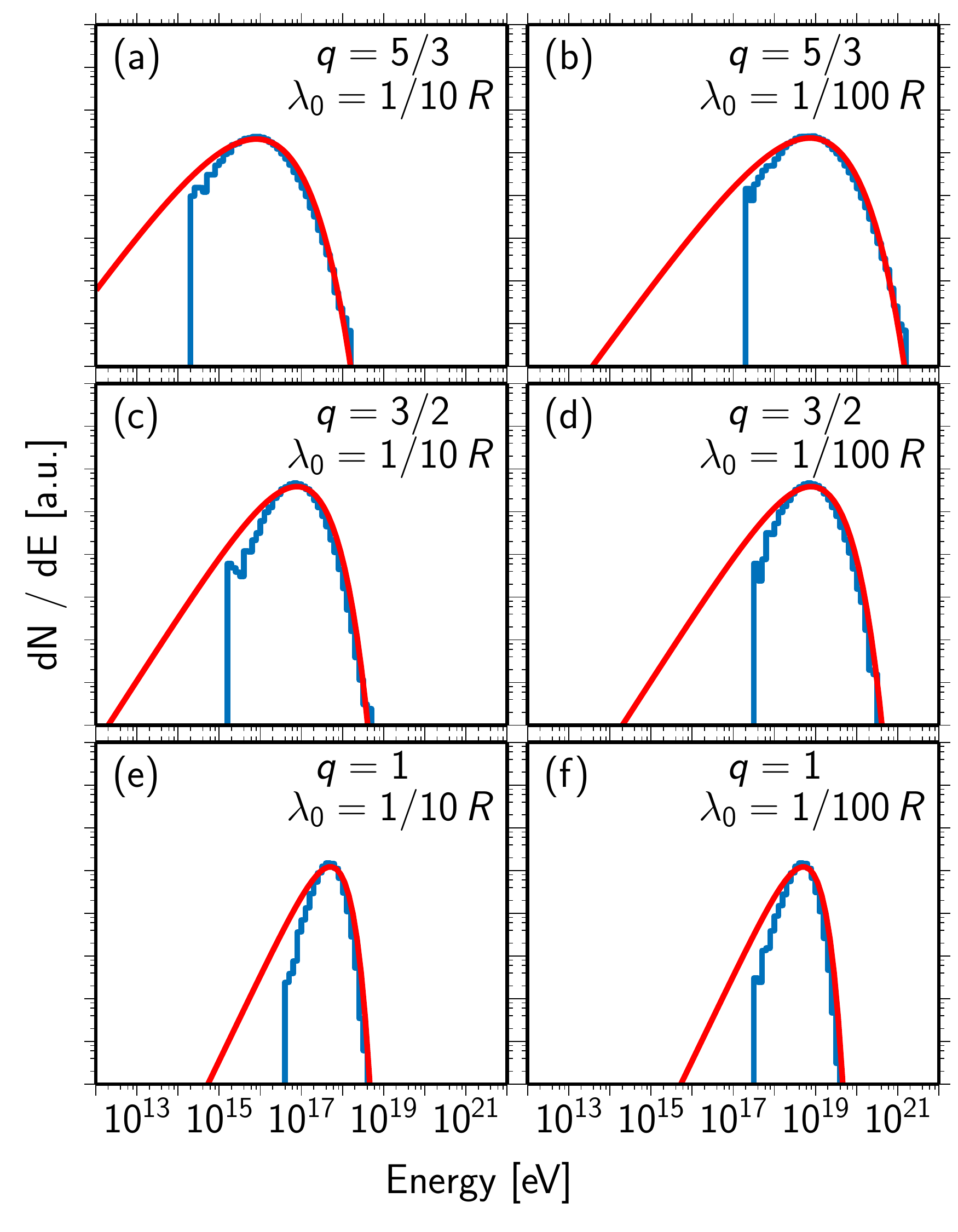}
	\caption{Emission spectra of the sources obtained from the simulation of second order Fermi acceleration
		of protons with different values for the mean free path length parameter
		$\lambda_0$ and the turbulence spectrum $q$ with $\beta = 0.05$. The red line shows a fit of
		eq.~\ref{eq:SourceSpectra} to the data.}
	\label{fig:AccelerationSpectra}
\end{figure}

In figure~\ref{fig:AccelerationSpectra} simulation results are shown for two
choices of $\lambda_0 / R$  and for three choices of $q$ covering a wide range
of scaling of the step length with energy.  In quasi linear theory these
correspond to Kolmogorov turbulence ($q =5/3$), Kraichnan turbulence ($q =
3/2$), and as an extreme case viscosity damping ($q = 1$)~\cite{Cho2003,
Cho2003a}. However, in realistic models of turbulence  the scaling parameter
$q$ can also assume different values depending on the plasma configuration.  In
particular, turbulence is anisotropic and $q = 5/3$ is for the perpendicular
spectrum only while for the parallel spectrum $q=2$~\cite{Beresnyak2015}. These
examples assume thus the effective step length as discussed above.

As the
particles are injected here in the center of the acceleration region, they
cannot escape without being accelerated first because of their small mean-free
path. This is different from e.g.\ shock acceleration where a fraction of
particles is lost after each cycle. In the model discussed here, low energy
particles can only escape from close to the border of the acceleration region,
while particles at the highest energies can escape from anywhere in the
accelerating region. This effectively suppresses the naively expected power-law
in the regime $\lambda_0 \ll R$, and the source spectrum can be reasonably well
be described by a peaking distribution
\begin{equation}
	\frac{dN}{dE} \propto E^{(3-q)}\, e^{- {(E/E_0)}^{(2-q)}}
	\label{eq:SourceSpectra}
\end{equation}
which we found on phenomenological grounds only.
For larger $\lambda_0$,
or $\lambda$ independent of the energy, both not shown here, the source spectra
develop a steep power-law tail. Here, we concentrate on the case where the
power-law tail is negligible small, respectively the energy range close to
$E_0$.

\begin{figure}[tb]
	\includegraphics[width=\columnwidth]{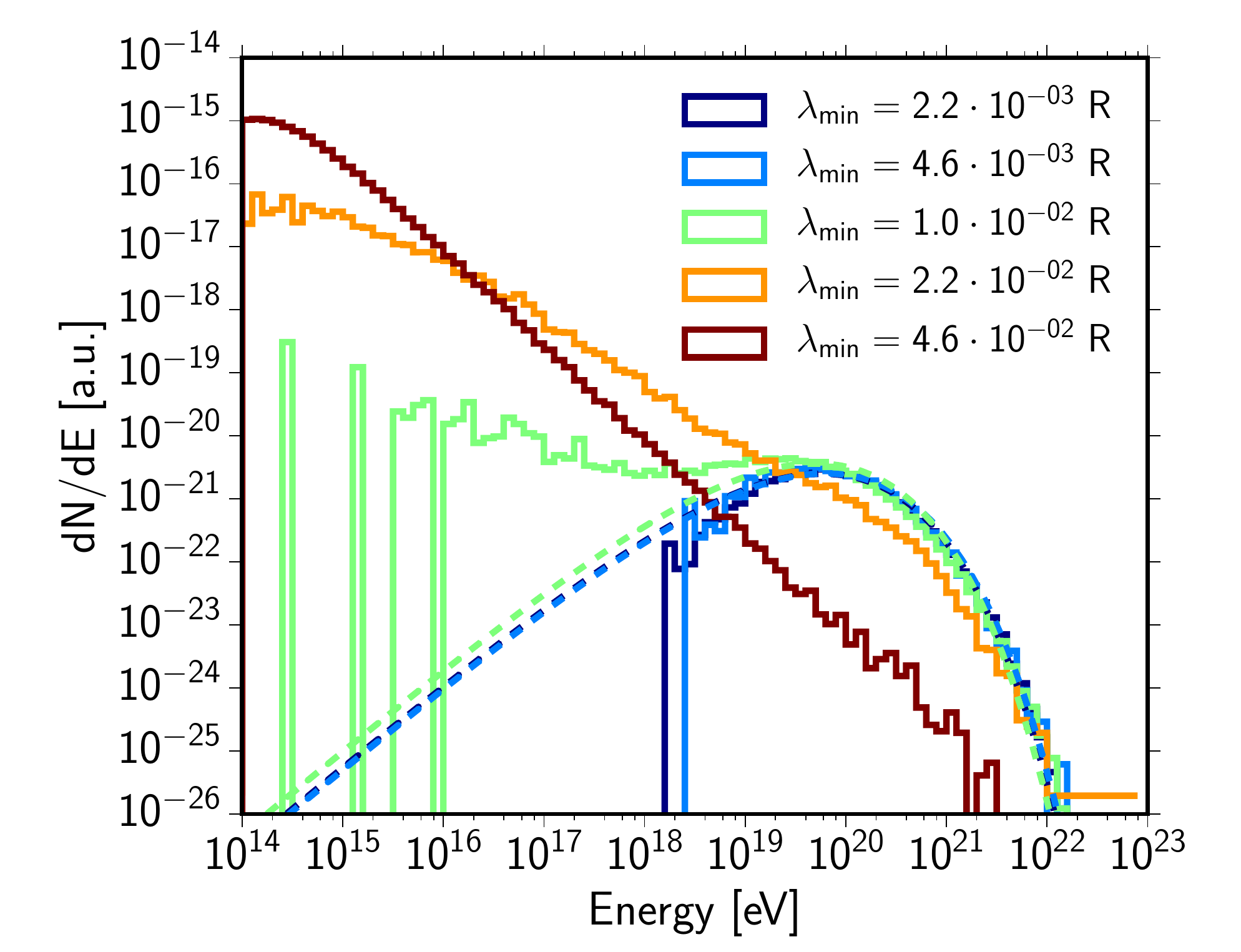}
	\caption{Source spectra with a fixed minimum step length $\lambda_{\min}$ for
	$q=5/3$, $\beta = 0.1$, and $\lambda_0 = \frac{1}{100}$ R. Dashed lines correspond to a fit of eq.~\ref{eq:SourceSpectra} to the simualted data above~\SI{30}{\EeV}.}
	\label{fig:Lambda_Min_Spectra}
\end{figure}
From detailed simulations of diffusive propagation in magnetic fields it is
known that a uniform scaling of the step length with a single power-law over the
whole energy range as considered here is unrealistic. As suggested by
simulations~\cite[e.g.][]{Beresnyak2011}, the step length is constant below a
threshold rigidity corresponding to a Larmour radius $r_L \approx \frac{1}{10} L_{\max}$
with turbulence injection scale $L_{\max}$. The source spectra obtained in
corresponding simulations with different choices of $\lambda_{\min}$
are shown in figure~\ref{fig:Lambda_Min_Spectra}. A minimum step length thus
has no effect on the obtained spectrum shape as long as the minimum step length
$\lambda_{\min}$ is still small compared to $R$ despite the transition. For
larger $\lambda_{\min}$ the spectrum transforms into a power-law with cut off
that can be still be described by the exponential in
eq.~\ref{eq:SourceSpectra}.

\begin{figure}[tb]
	\includegraphics[width=\columnwidth]{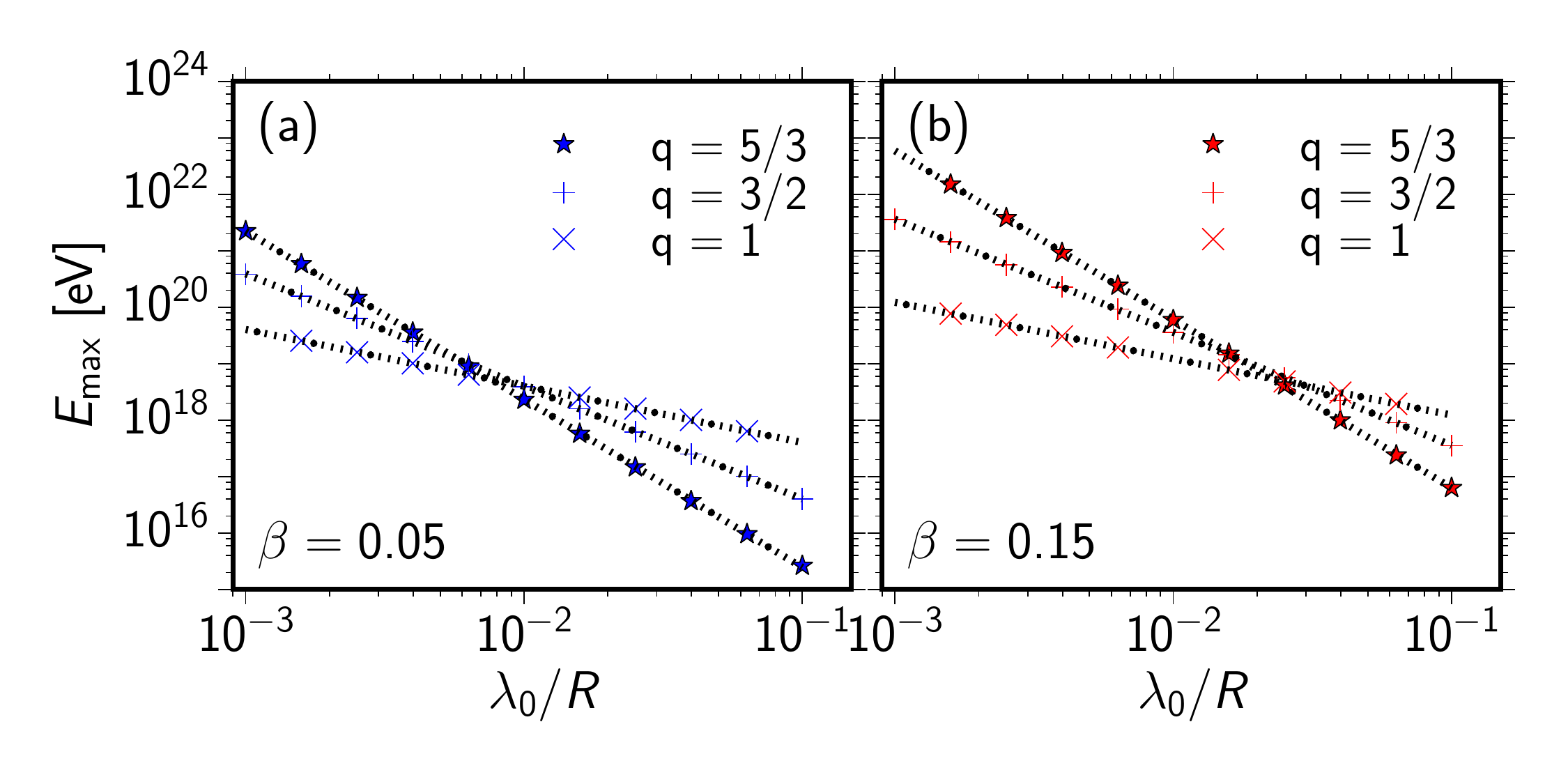}
	\caption{Maximum of the source spectra $E_{\max}$ as function of the
		scatter length $\lambda_0$ for three choices of the turbulence spectrum $q$
		for \textbf{(a)} $\beta = 0.05$ and \textbf{(a)} $\beta = 0.15$. Colored
		markers are from a fit of eq.~\ref{eq:SourceSpectra} to simulations as
		e.g.\ shown in figure~\ref{fig:AccelerationSpectra}; the dashed
	lines correspond to a power-law fit to the simulated points. }
	\label{fig:SourceDistributionLocation}
\end{figure}
For a given size of the accelerating region, the parameter $E_0$ of the
distribution depends on the mean free path length $\lambda_0$, the velocity of
the scatter centers $\beta$, and the spectral index $q$ of the magnetic
turbulence.    From
eq.~\ref{eq:SourceSpectra} follows, that the position of the maximum of the
distribution is given by
\begin{equation}
	E_{\max} = {\left(\frac{3-q}{2-q}\right)}^{\frac{1}{2-q}}\; E_0.
	\label{}
\end{equation}
The dependency of $E_{\max}$ on $\lambda_0$ is shown in
figure~\ref{fig:SourceDistributionLocation}.  We found no dependency of the
shape of the spectrum or the maximum energy on the injection energy. The
spectral shape is thus independent on the injection spectrum and thus the
properties of the pre-accelerator as long as the energies reached by the
pre-accelerator are not  too high.  A fit of a power-law $E_{\max}
\propto {\lambda_0}^{\left(\frac{1}{q-2}\right)}$ to the simulation perfectly
describes the dependency.  This is consistent with eq.~\ref{eq:MeanFreePath},
as this relation implies a linear dependency of $E_{\max}$ on the charge $Z$ of
the particles.

\begin{figure}[tb]
	\includegraphics[width=\columnwidth]{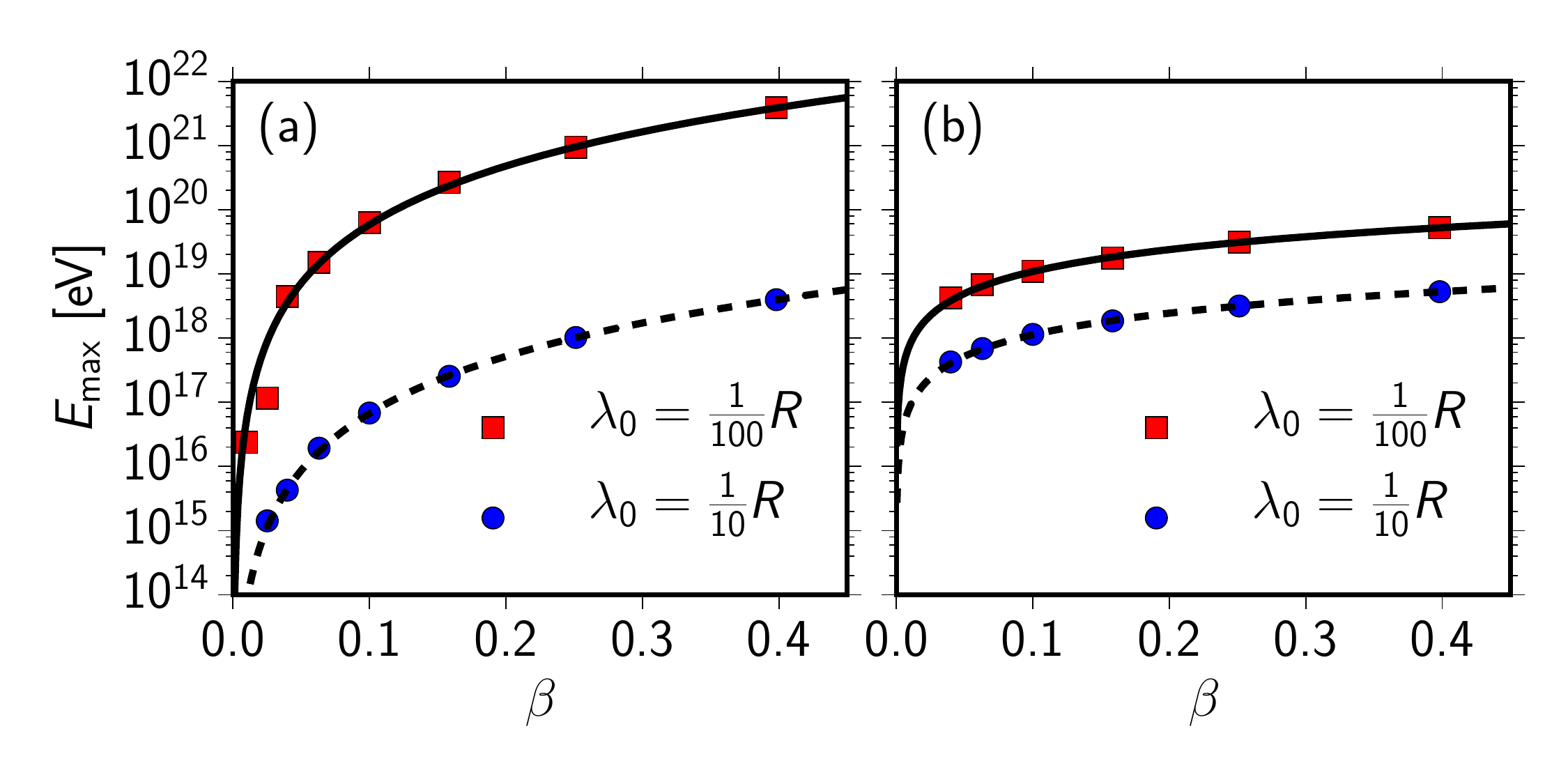}
	\caption{Maximum of the source spectra $E_{\max}$ as function of the
	velocity of the scatter centers $\beta$ for two choices of $\lambda_0$ for \textbf{(a)} $q = 5/3$ and \textbf{(b)} $q = 1$. Dashed and dotted lines correspond to a fit of $E_{\max} \propto \beta^a$. }
	\label{fig:maxEnergyBetaDependency}
\end{figure}

In figure~\ref{fig:maxEnergyBetaDependency} the dependency of $E_{\max}$ on
$\beta$ is shown for two choices of $\lambda_0$ for $q=5/3$ and $q=1$. Points corresponds to a fit of eq.~\ref{eq:SourceSpectra} to the simulation result and lines to a fit of $E_{\max} \propto \beta^a$ to the corresponding simulation sets. For $q=5/3$  we found $a = 3.03 \pm 0.02$ and for $q = 1$ we found $a = 1.12 \pm 0.02$.

\section{Fit}
\begin{figure}[tb]
	\includegraphics[width=\columnwidth]{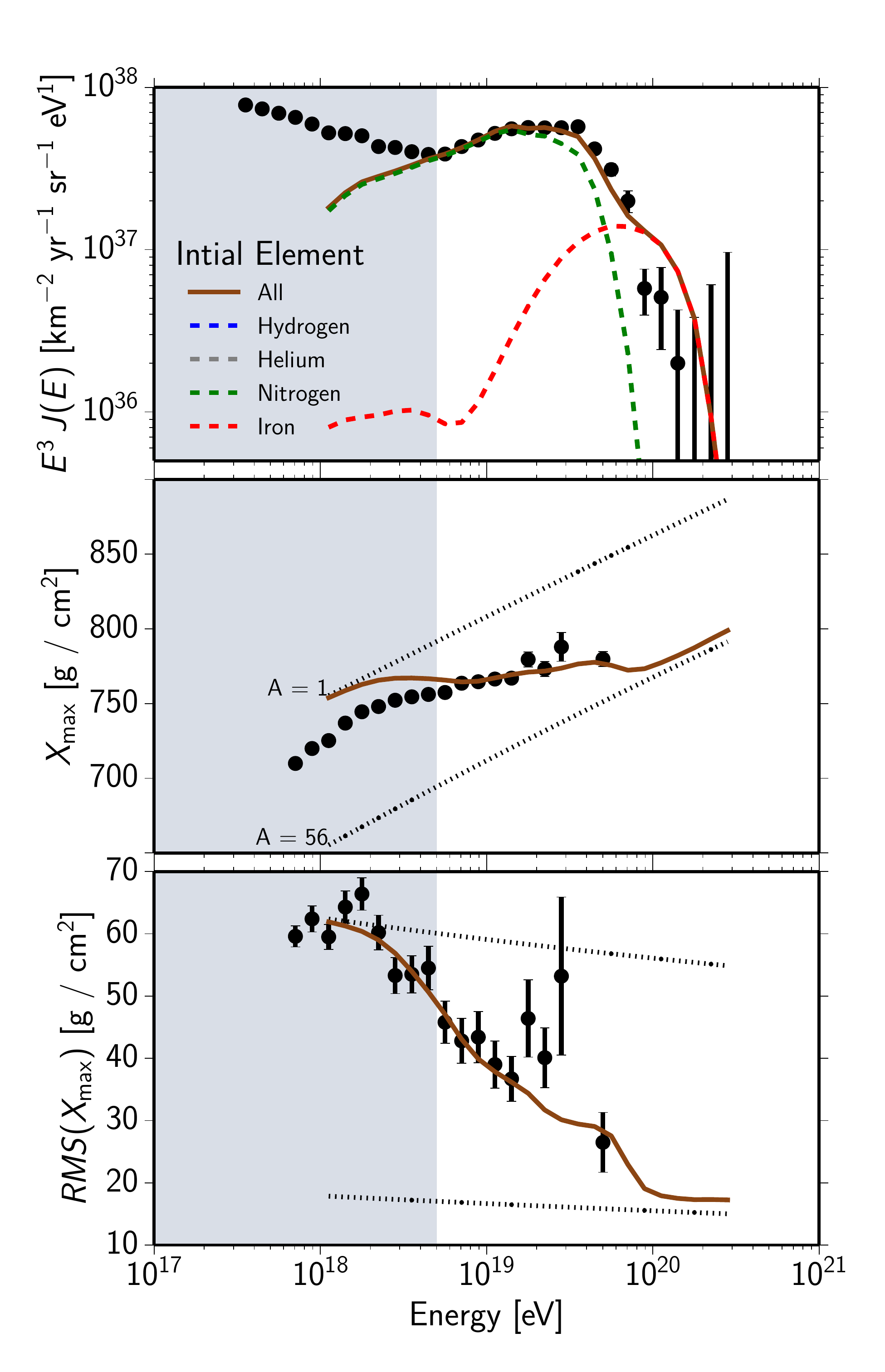}
	\caption{Two parameter fit of the model described here to the data of the
		energy spectrum~\cite{PAO2015} (top), and the mean (middle) and root-mean-square
		(bottom) of the  $X_{\max}$ distributions measured with the Pierre Auger
		Observatory~\cite{Aab2014f}. For the spectrum fit the contribution from the
	individual initial elements are  marked as dashed lines.  The gray marked
area is not included in the fit.}
	\label{fig:ObservableDistributions}
\end{figure}
We fit the source spectrum given by eq.~\ref{eq:SourceSpectra} to the
observational data collected by the Pierre Auger Observatory~\cite{PAO2015a}
accounting for propagation effects using simulations of UHECR propagation with
the CRPropa software~\cite{Batista2016}.  We use the data on the flux of
ultra-high energy cosmic rays~\cite{PAO2015} and the depth of shower maximum,
$X_{\max}$, i.e.~the depth at which the energy deposit of the electromagnetic
cascade reaches its maximum~\cite{Aab2014f}.
The 1D-simulations used here include energy losses and production of
secondaries from  pion-production, photo-disintegration, and electron pair
production in the cosmic microwave background and infrared background of
particles from sources up to $\SI{3}{\giga\parsec}$.  Energy losses by
adiabatic expansion of the universe and energy losses and secondary production
by nuclear decay are included as well.  As a model for the infrared background
we consider here only Gilmore2012~\cite{Gilmore2012} and as model for the
nuclear cross-sections only TALYS~\cite{Koning2005}. An overview of the effects
of these models on fits to cosmic ray data and simulation uncertainties is
given in reference~\cite{AlvesBatista2015a}.

The fit-method used here is based on the method developed for the
interpretation of the data of the Pierre Auger Observatory with power-law
source spectra and an exponential cut-off~\cite{Matteo2015, Walz2016, Aab2016l}.
To evaluate the effect of a source model on the observed cosmic rays, the
energy and mass binned cosmic rays of a propagation simulation are re-weighted
according to the source parameters. From the resulting weights for observed
cosmic rays with energy and mass ${(E, A)}_i$ the spectrum and $X_{\max}$
distributions are calculated. For the calculation of $X_{\max}$ from ${(E,
A)}_i$ we use the usual parametrization based on the Gumble distribution and
EPOS-LHC~\cite{Werner2011} as the hadronic interaction
model~\cite{DeDomenico2013a}. The optimum set of source parameters is obtained
by maximizing the likelihood of the observations using a Markov chain Monte
Carlo code~\cite{Patil2010}. Differently from the original approach, here we
use a Gaussian-based likelihood for the mean and RMS of the  $X_{\max}$
distribution.

The best fit  is obtained for $\log_{10}{(E_{\max} / \text{eV} / Z)} = 18.21 \pm
0.02$ and $q = 1.29 \pm 0.02$. With this fitting procedure, individual Markov
Chains converge to different values for the individual element fractions while
the values for $q$ and $E_{\max}$ are stable as indicated by a good
Gelman-Rubin~\cite{Gelman1992} statistic. The individual chains converge to
different values for the individual elements, but the ratio of the Iron to
Nitrogen fraction is well constrained in all Markov Chains to $f_\text{Fe} /
f_\text{N} = (5.6 \pm 0.7)\cdot10^{-4}$, and the values for Hydrogen and Helium
are almost arbitrary due to the low maximum rigidity obtained in the fit. With
these parameters, the expected values of the observables  are shown together
with the data in figure~\ref{fig:ObservableDistributions}.  For this four
parameter fit to 36 data points we obtain a goodness-of-fit $\chi^2 /\text{dof}
= 58.8 / 32$.

\section{Discussion}
\subsection{Fit}
The model presented here fits the data approximately as well as fits assuming
power-laws~\cite{Matteo2015, Aab2016l}. This is not surprising, as from a mere technical
point of view the spectrum proposed here is an inverted power-law with spectral
index limited to $ -2 \leq \gamma \leq -\frac{4}{3}$ and a slightly modified
exponential cut off; results with $\gamma \le 0$ have already been reported
from fits of power-laws~\cite{Matteo2015}.  However, in the scenario as
discussed here, this previously surprising parameter range is a natural
consequence of the acceleration mechanism. The particular shape of the cut-off
reduces the degeneracy between the parameters faced in fitting conventional
power-law models. In particular, the cut-off expected here hardens the end of
the spectrum compared to a naive exponential or broken-exponential cut-off;
only for $q=1$ the shape of the cut-off is identical to the common exponential
cut-off. Consequently, the data above \SI{5}{\EeV} can be described with heavy
primaries only and the fit becomes insensitive to the abundances of lighter
elements.

Here we have considered only four elements, Hydrogen, Helium, Nitrogen, and
Iron, where Nitrogen acts as a proxy for all elements with $2 < Z < 26$.
Including more individual heavy elements in the fit will reduce the parameter
$E_{\max}$ further, as it allows a better description of the cut-off consistent
with the `disappointing model'~\cite{Aloisio2011}.  The sources of
extragalactic cosmic rays thus do not need to be surprisingly metal-rich as
implied by power-law fits~\cite{Boncioli2015}.  One can further speculate, that
thus in this model the extragalactic sources of UHECR are also the origin of
the strong light component below the ankle that have been recently
measured~\cite{Porcelli2015, Buitink2016}. In addition to the inclusion of more
elements, a detailed investigation of this requires the extension of the fit
regime to lower energies and the inclusion of Galactic models in the fashion
of~\cite{Thoudam2016} and is beyond the scope of this paper.

\subsection{Simulation}
The approximation of diffusion in our simulation and the corresponding
parametrizations are motivated by quasi-linear theory. QLT is based on the regime
where the gyro-radius of the particle is not larger than the
coherence length $\Lambda$ of the field~\cite{Schlickeiser1989, Casse2001}.
The correlation length $\Lambda$ of a turbulent magnetic field with energy
distribution following a power-law with index $q$ between length-scales
$L_{\min}$ and $L_{\max}$ is defined as $\Lambda = \frac{1}{2} L_{\max}
\frac{q-1}{q} \frac{(1 - L_{\min} /
L_{\max})^q}{1-(L_{\min})/L_{\max})^{q-1}}$.  For $q > 1$ and $L_{\min} /
L_{\max} \rightarrow 0$ we get $\Lambda = \frac{1}{2} L_{\max} \frac{q-1}{q}$;
For $q \rightarrow 1 $ we get a lower bound for the coherence length $\Lambda >
\frac{1}{2} L_{\max} \frac{1-L_{\min} / L_{\max}}{-\ln L_{\min} / L_{\max}}$ as
function of the span of the turbulence scales.  Assuming that the maximum
turbulence scale is of the size of the accelerating region yields that even for
the extreme case $q = 1$ quasi-linear theory is applicable if the turbulence
scales spans only two orders of magnitude and the accelerator is approximately
ten-times larger than the gyro-radius.  The constraints on $L_{\min} /
L_{\max}$ are greatly relaxed for $q = 5/3$, $q=3/2$, or $q=1.3$ corresponding
to the best fit to the Auger data as obtained in this work.
However, QLT does not account for several scattering processes likely found in
realistic plasmas and may dominate particle propagation~\cite[e.g.][]{Yan2002, Lazarian2014}.
Nevertheless, an increase of the step length parametrised by a power law as
used in our simulation also expected in at least some non-linear
theories~\cite{Shalchi2009} and seen in simulations~\cite{Beresnyak2011}, at
least above a certain threshold rigidity.  Quantitative conclusions from the
fit-result on the properties of the accelerator are thus not straight forward
and require much ore detailed modelling of the acceleration environment.

By using a test particle approach for the simulations, we implicitly assumed
that only a small portion of energy stored in the turbulent field is transferred
to UHECR, because otherwise the turbulence spectrum is modified by
collision-less damping, which would alter the spectrum of cosmic rays.  However, as the
spectrum from our acceleration mechanism is peaking and not a wide power-law,
the energetics are here not dominated by the low-energy cosmic rays.  This
reduces the energetic constraints in this mechanism compared to mechanisms
predicting a power-law. To obtain a conservative estimate, we attribute the
UHECR flux to originate only from a single source in $ D
=\SI{50}{\mega\parsec}$ distance, yielding a cosmic ray luminosity of $ L_{CR}
\approx \SI{6d42}{\erg\per\second}$ for the best fit parameters and assuming
iron primaries only.  Thus even if  only a single source accounts for the  flux
of UHECR, the energy transferred from the magnetic field to cosmic rays is much
less then the energy transferred to radio emission in e.g.\ AGNs as sources. In
case of multiple sources, the energy transferred to UHECR in a single source is
much smaller.
\begin{figure*}[tb]
	\centering
	\includegraphics[width=.7\textwidth]{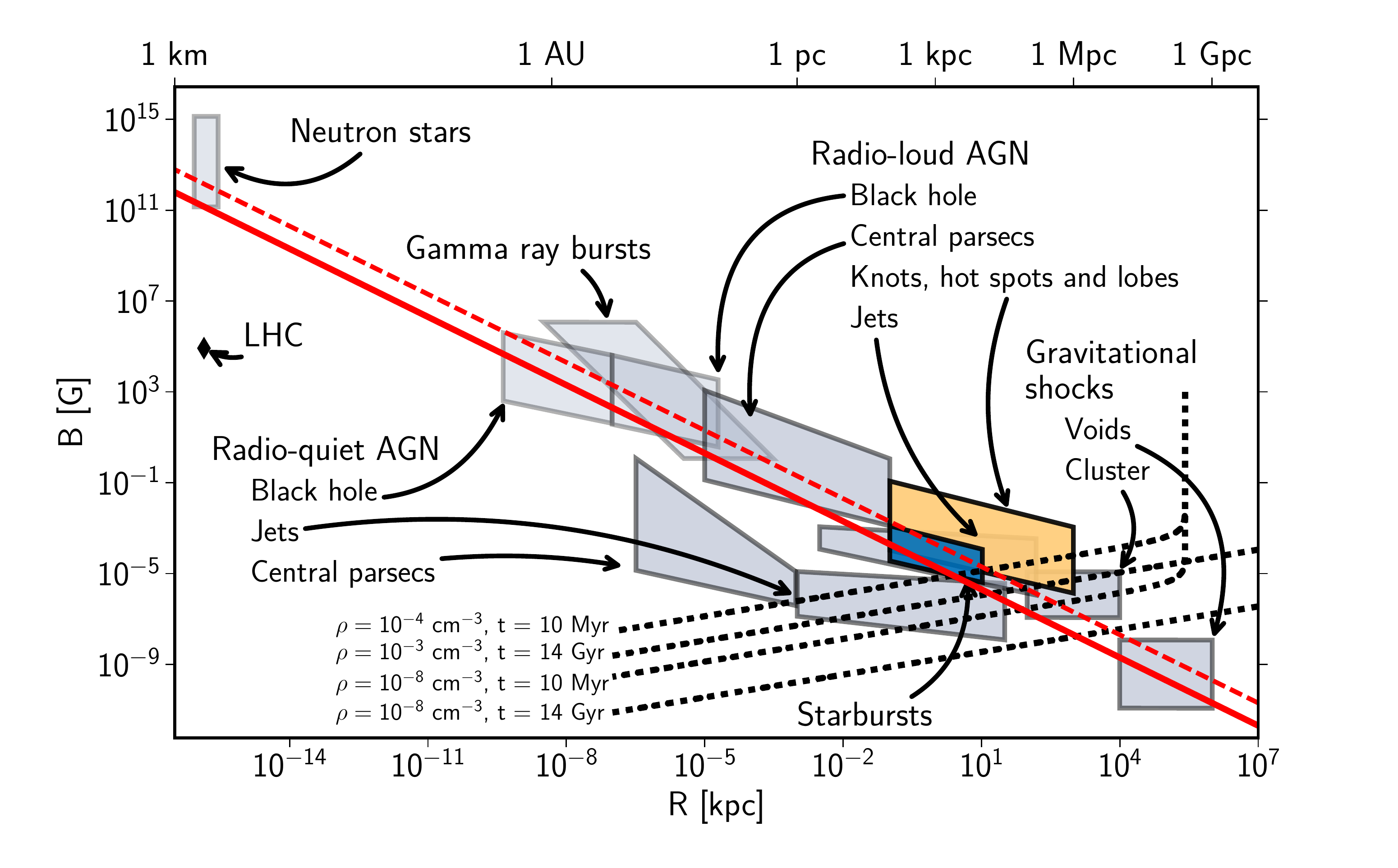}
	\caption{Possible accelerators for UHECR following an idea of
		Hillas'~\cite{Hillas1984}, based on data published in
		reference~\cite{Ptitsyna2010}. The solid
		red line marks the gyro-radius $R_G$ of an iron particle with
		\SI{260}{\EeV}, the red-dashed line $10 R_G$.
		The black dotted lines correspond to time constraints following
	eq.~\ref{eq:accelerationtime} with maximum time $t$ and density $\rho$ as
indicated left of the  respective lines. The parameterspace below the
respective lines is excluded.}
	\label{fig:HillasPlot}
\end{figure*}

Within the simulations we assume a constant, scale independent velocity of the
scatter centers. However, due to energy cascading the velocity of the scatter
centers may be reduced at low energies reducing the efficiency of this
acceleration mechanism and thus sets an additional requirement to the
pre-accelerator. Furthermore, we neglected any energy losses in particular by
interaction with background photon fields in the source environments, as our
simulations do not account correctly for the acceleration time due to the
simplistic parametrization of the diffusion tensor.

\subsection{Source Candidates}
Following an idea of Hillas', potential sources can be discussed
in terms of the maximum achievable energy given their size and magnetic field strength.
The source spectrum obtained here contains a natural cut-off beyond $E_{\max}$.
However, the spectrum cannot continue indefinitely with exponential decay only
as an absolute limiting energy $E_{\lim}$ is given by the largest possible
gyro-radius of the particles in the respective magnetic field. Here we can
safely use $E_{\lim} \approx 10 E_{\max} $ or $E_{\lim} = \SI{260}{\EeV}$ for
Iron nuclei consistent with observations.  The classical constraint that the
size of the accelerator has to be larger than the gyro-radius of the
particles~\cite{Hillas1984} provides thus not a severe constraint here and
leaves in particular radio-loud AGN, starbursts and turbulence induced by
gravitational shocks~\cite{Ptitsyna2010} as possible sources with this
acceleration mechanism as indicated in figure~\ref{fig:HillasPlot}.

The main argument against second order Fermi acceleration is commonly the required acceleration time.
A rough upper
limit for the acceleration time with second order Fermi acceleration is
\begin{equation} t_{\text{acc}} = 50 \left(\frac{\beta}{\SI{7d-4}{}}
	\right)^{-2} \left( \frac{R}{\SI{10}{\kilo\parsec}}
	\right)^{\frac{2}{3}}\left(\frac{E}{\SI{d19}{\electronvolt}}
	\right)^{\frac{1}{3}}\text{Gyr},
	\label{eq:accelerationtime}
\end{equation}
assuming again that the maximum length-scale of the turbulence is the size of
the accelerator~\cite{OSullivan2009}. As via the Alfv\'en velocity  $\beta
\propto \frac{B}{\sqrt{\rho_P}}$ the acceleration time depends on the strength
of the magnetic field $B$, eq.~\ref{eq:accelerationtime} can be used to
constrain possible accelerators in a Hillas' plot given their lifetime and
plasma density~$\rho_P$.
In general, the acceleration mechanism thus selects large low-density environments.
The largest low-density environments for considered UHECR acceleration are plasmas in clusters and voids
with turbulence induced by gravitational shocks.
Assuming a lifetime of~\SI{14}{\giga\years} and densities
of~\SI{d-3}{\per\cubic\centi\meter} as inter-cluster density respectively
\SI{d-8}{\per\cubic\centi\meter} for voids as extreme values~\cite{Peterson2006}, we can exclude
turbulences induced by gravitational shocks as candidates for this mechanism.
Radio lobes of AGNs as well as superwinds of starburst galaxies
have a lifetime of more than \SI{d7}{\years}. If we assume a density
below~\SI{d-4}{\per\cubic\centi\meter} for the jets and lobes of radio-loud
AGNs as well as superwinds emitted by starburst galaxies, both
classes remain possible acceleration sites for this scenario as indicated in
figure~\ref{fig:HillasPlot}. This is consistent with similar estimates by
Hillas~\cite{Hillas1984}.

Here of course we require that the turbulence spectrum covers large enough
scales for the acceleration process. The highest energies can thus not be
reached by this mechanism in the inner parts of AGN. Also, if the turbulence
scale in the superwind of starbursts is much lower than their size,
corresponding e.g.\ a ten times tighter constraint on the gyro radius, only few
starburst galaxies would provide the required conditions.  However, this is no
problem for potential acceleration in the radiolobes of AGN.

Both remaining candidate classes also fulfill the luminosity
requirement~\cite{Lemoine2009, Hardcastle2009}, that for any UHECR source  $L>
\SI{d45}{}\left(\frac{E / Z}{\SI{d20}{\electronvolt}}\right)^{2}
\SI{}{\erg\per\second}$, which for the iron nuclei as above reads $L >
\SI{1d43}{\erg\per\second}$.
With abundances of about
\SI{d-4}{\per\cubic\mega\parsec} both classes are furthermore consistent with
observational bounds on UHECR~\cite{Abreu2013a, Aab2015f} and neutrino
sources~\cite{Aartsen2016} as resulting from the low levels of anisotropy in
the respective arrival directions.

\section{Conclusion}
The acceleration mechanism outlined here suggests the following picture as a
potential scenario for the origin of UHECR. Cosmic-ray particles are first
pre-accelerated in the inner parts of astronomical objects as potentially AGNs
or starburst Galaxies (c.f.~\cite{Anchordoqui1999}) to an intermediate energy
above the dissipation scale of the surrounding magnetic field in the AGN
radiolobes respectively superwind of the starburst galaxies. In a second step
the highest energies are then reached via the second order Fermi mechanism
resulting in a peaking emission spectrum and not the commonly expected
power-law. In contrast to interpretations of the data with power-laws as
produced by shock acceleration, the best fitting parameters of the peaking
source spectrum proposed here are not exceptional and do not imply an
unexpected high abundance of heavy elements at the sources.  It may thus be
worth to reconsider the role of the usually disfavoured second-order Fermi
mechanism for the acceleration of UHECR.

%
%
%
\appendix

\section*{Acknowledgements}
	We gratefully acknowledge valuable comments to the manuscript from Roger
	Clay, Martin Erdmann, David Walz, and three anonymous referees. This research
	was funded by the European Research Council (ERC) under the European Union's
	Horizon 2020 research and innovation programme (grant agreement No 640130).

\section*{References}
\bibliographystyle{elsarticle-num}
\bibliography{bibliography}

\begin{thebibliography}{10}
\expandafter\ifx\csname url\endcsname\relax
  \def\url#1{\texttt{#1}}\fi
\expandafter\ifx\csname urlprefix\endcsname\relax\def\urlprefix{URL }\fi
\expandafter\ifx\csname href\endcsname\relax
  \def\href#1#2{#2} \def\path#1{#1}\fi

\bibitem{Kampert2014}
K.-H. Kampert, P.~Tinyakov, {Cosmic rays from the ankle to the cutoff}, Comptes
  Rendus Physique 15 (2014) 318--328.
\newblock \href {http://arxiv.org/abs/1405.0575} {\path{arXiv:1405.0575}},
  \href {http://dx.doi.org/10.1016/j.crhy.2014.04.006}
  {\path{doi:10.1016/j.crhy.2014.04.006}}.

\bibitem{Kotera2011}
K.~Kotera, A.~V. Olinto, The astrophysics of ultrahigh energy cosmic rays,
  Annual Review of Astronomy and Astrophysics 49 (2011) 119--153.
\newblock \href {http://arxiv.org/abs/1101.4256} {\path{arXiv:1101.4256}},
  \href {http://dx.doi.org/10.1146/annurev-astro-081710-102620}
  {\path{doi:10.1146/annurev-astro-081710-102620}}.

\bibitem{Colgate1994}
S.~Colgate, {Acceleration in astrophysics}, Physica Scripta T52 (1994) 96--105.
\newblock \href {http://dx.doi.org/10.1088/0031-8949/1994/T52/017}
  {\path{doi:10.1088/0031-8949/1994/T52/017}}.

\bibitem{Blasi2012}
P.~Blasi, Theoretical challenges in acceleration and transport of ultra high
  energy cosmic rays: A review, Invited Review Talk in UHECR2012 Symposium,
  CERN\href {http://arxiv.org/abs/1208.1682} {\path{arXiv:1208.1682}}.

\bibitem{Marcowith2016}
A.~Marcowith, A.~Bret, A.~Bykov, M.~E. Dieckman, L.~O. Drury, B.~Lembège,
  M.~Lemoine, G.~Morlino, G.~Murphy, G.~Pelletier, I.~Plotnikov, B.~Reville,
  M.~Riquelme, L.~Sironi, A.~S. Novo, The microphysics of collisionless shock
  waves, Reports on Progress in Physics 79~(4) (2016) 046901.

\bibitem{Matteo2015}
{A. di Matteo for the Pierre Auger Collaboration}, Combined fit of spectrum and
  composition data as measured by the pierre auger observatory, in: Proceedings
  of the 34th International Cosmic Ray Conference, 2015.

\bibitem{Aab2016l}
A.~Aab, et~al., {Combined fit of spectrum and composition data as measured by
  the Pierre Auger Observatory}, JCAP 1704~(04) (2017) 038.
\newblock \href {http://arxiv.org/abs/1612.07155} {\path{arXiv:1612.07155}},
  \href {http://dx.doi.org/10.1088/1475-7516/2017/04/038}
  {\path{doi:10.1088/1475-7516/2017/04/038}}.

\bibitem{Allard2008}
D.~Allard, N.~Busca, G.~Decerprit, A.~V. Olinto, E.~Parizot, Implications of
  the cosmic ray spectrum for the mass composition at the highest energies,
  Journal of Cosmology and Astroparticle Physics 10 (2008) 033.
\newblock \href {http://arxiv.org/abs/0805.4779} {\path{arXiv:0805.4779}}.

\bibitem{Aloisio2014}
R.~Aloisio, V.~Berezinsky, P.~Blasi, Ultra high energy cosmic rays:
  implications of auger data for source spectra and chemical composition,
  Journal of Cosmology and Astroparticle Physics 2014~(10) (2014) 020.

\bibitem{Unger2015}
M.~Unger, G.~R. Farrar, L.~A. Anchordoqui, {Origin of the ankle in the
  ultrahigh energy cosmic ray spectrum, and of the extragalactic protons below
  it}, Phys. Rev. D92~(12) (2015) 123001.
\newblock \href {http://arxiv.org/abs/1505.02153} {\path{arXiv:1505.02153}},
  \href {http://dx.doi.org/10.1103/PhysRevD.92.123001}
  {\path{doi:10.1103/PhysRevD.92.123001}}.

\bibitem{Fermi1949}
E.~Fermi, On the origin of the cosmic radiation, Physical Review 75 (1949)
  1169.

\bibitem{Pelletier1999}
G.~{Pelletier}, {Cosmic ray acceleration and nonlinear relativistic
  wavefronts}, Astronomy and Astrophysics 350 (1999) 705--718.

\bibitem{OSullivan2009}
S.~{O'Sullivan}, B.~{Reville}, A.~M. {Taylor}, {Stochastic particle
  acceleration in the lobes of giant radio galaxies}, Monthly Notices of the
  Royal Astronomical Society 400 (2009) 248--257.
\newblock \href {http://arxiv.org/abs/0903.1259} {\path{arXiv:0903.1259}},
  \href {http://dx.doi.org/10.1111/j.1365-2966.2009.15442.x}
  {\path{doi:10.1111/j.1365-2966.2009.15442.x}}.

\bibitem{Fraschetti2008}
F.~Fraschetti, On the acceleration of ultra-high-energy cosmic rays,
  Philosophical Transactions of the Royal Society A: Mathematical, Physical and
  Engineering Sciences 366~(1884) (2008) 4417--4428.
\newblock \href {http://arxiv.org/abs/0809.3057} {\path{arXiv:0809.3057}},
  \href {http://dx.doi.org/10.1098/rsta.2008.0204}
  {\path{doi:10.1098/rsta.2008.0204}}.

\bibitem{Hardcastle2009}
M.~J. Hardcastle, C.~C. Cheung, I.~J. Feain, L.~Stawarz, {High-energy Particle
  Acceleration and Production of Ultra-high-energy Cosmic Rays in the Giant
  Lobes of Centaurus A}, Monthly Notices of the Royal Astronomical Society 393
  (2009) 1041--1053.
\newblock \href {http://arxiv.org/abs/0808.1593} {\path{arXiv:0808.1593}},
  \href {http://dx.doi.org/10.1111/j.1365-2966.2008.14265.x}
  {\path{doi:10.1111/j.1365-2966.2008.14265.x}}.

\bibitem{Lazarian2012}
A.~{Lazarian}, L.~{Vlahos}, G.~{Kowal}, H.~{Yan}, A.~{Beresnyak}, E.~M. {de
  Gouveia Dal Pino}, {Turbulence, Magnetic Reconnection in Turbulent Fluids and
  Energetic Particle Acceleration}, Space Science Reviews 173 (2012) 557--622.
\newblock \href {http://arxiv.org/abs/1211.0008} {\path{arXiv:1211.0008}},
  \href {http://dx.doi.org/10.1007/s11214-012-9936-7}
  {\path{doi:10.1007/s11214-012-9936-7}}.

\bibitem{Schlickeiser1989}
R.~{Schlickeiser}, {Cosmic-Ray Transport and Acceleration. II. Cosmic Rays in
  Moving Cold Media with Application to Diffusive Shock Wave Acceleration}, The
  Astrophysical Journal 336 (1989) 264.
\newblock \href {http://dx.doi.org/10.1086/167010} {\path{doi:10.1086/167010}}.

\bibitem{Scalo2004}
J.~Scalo, B.~G. Elmegreen, Interstellar turbulence ii: Implications and
  effects, Annual Review of Astronomy and Astrophysics 42~(1) (2004) 275--316.
\newblock \href {http://dx.doi.org/10.1146/annurev.astro.42.120403.143327}
  {\path{doi:10.1146/annurev.astro.42.120403.143327}}.

\bibitem{Casse2001}
F.~Casse, M.~Lemoine, Pelletier, Transport of cosmic rays in chaotic magnetic
  fields 65.

\bibitem{Goldreich1995}
P.~{Goldreich}, S.~{Sridhar}, {Toward a theory of interstellar turbulence. 2:
  Strong alfvenic turbulence}, Astrophysical Journal 438 (1995) 763--775.
\newblock \href {http://dx.doi.org/10.1086/175121} {\path{doi:10.1086/175121}}.

\bibitem{Yan2002}
H.~Yan, A.~Lazarian, {Scattering of cosmic rays by magnetohydrodynamic
  interstellar turbulence}, Phys. Rev. Lett. 89 (2002) 281102.
\newblock \href {http://arxiv.org/abs/astro-ph/0205285}
  {\path{arXiv:astro-ph/0205285}}, \href
  {http://dx.doi.org/10.1103/PhysRevLett.89.281102}
  {\path{doi:10.1103/PhysRevLett.89.281102}}.

\bibitem{Lazarian2014}
A.~Lazarian, H.~Yan, {Superdiffusion of Cosmic Rays: Implications for Cosmic
  Ray Acceleration}, Astrophys. J. 784 (2014) 38.
\newblock \href {http://arxiv.org/abs/1308.3244} {\path{arXiv:1308.3244}},
  \href {http://dx.doi.org/10.1088/0004-637X/784/1/38}
  {\path{doi:10.1088/0004-637X/784/1/38}}.

\bibitem{Yan2004}
H.~Yan, A.~Lazarian, {Cosmic ray scattering and streaming in compressible
  magnetohydrodynamic turbulence}, Astrophys. J. 614 (2004) 757--769.
\newblock \href {http://arxiv.org/abs/astro-ph/0408172}
  {\path{arXiv:astro-ph/0408172}}, \href {http://dx.doi.org/10.1086/423733}
  {\path{doi:10.1086/423733}}.

\bibitem{Shalchi2009}
A.~Shalchi, Nonlinear Cosmic Ray Diffusion Theories, Springer, 2009.
\newblock \href {http://dx.doi.org/10.1007/978-3-642-00309-7}
  {\path{doi:10.1007/978-3-642-00309-7}}.

\bibitem{Beresnyak2011}
A.~Beresnyak, H.~Yan, A.~Lazarian, Numerical study of cosmic ray diffusion in
  magnetohydrodynamic turbulence, The Astrophysical Journal 728~(1) (2011) 60.

\bibitem{Batista2016}
R.~{Alves Batista}, et~al., Crpropa 3 - a public astrophysical simulation
  framework for propagating extraterrestrial ultra-high energy particles 05
  (2016) 038.
\newblock \href {http://arxiv.org/abs/1603.07142} {\path{arXiv:1603.07142}}.

\bibitem{Cho2003}
J.~Cho, A.~Lazarian, E.~T. Vishniac, Ordinary and viscosity-damped
  magnetohydrodynamic turbulence, The Astrophysical Journal 595~(2) (2003) 812.

\bibitem{Cho2003a}
J.~Cho, A.~Lazarian, E.~Vishniac, {MHD Turbulence: Scaling Laws and
  Astrophysical Implications}, 2003, pp. 56--100.
\newblock \href {http://arxiv.org/abs/astro-ph/0205286}
  {\path{arXiv:astro-ph/0205286}}.

\bibitem{Beresnyak2015}
A.~{Beresnyak}, {On the Parallel Spectrum in Magnetohydrodynamic Turbulence},
  The Astrophysical Journal Letters 801 (2015) L9.
\newblock \href {http://arxiv.org/abs/1407.2613} {\path{arXiv:1407.2613}},
  \href {http://dx.doi.org/10.1088/2041-8205/801/1/L9}
  {\path{doi:10.1088/2041-8205/801/1/L9}}.

\bibitem{PAO2015}
I.~V. for~the Pierre Auger~Collaboration, The flux of ultra-high energy cosmic
  rays after ten years of operation of the pierre auger observatory, in:
  Proceedings of the 34th International Cosmic Ray Conference, 2015.

\bibitem{Aab2014f}
A.~Aab, et~al., {Depth of maximum of air-shower profiles at the {Pierre Auger
  Observatory}. {I. Measurements} at energies above $10^{17.8}$ eV}, Phys. Rev.
  D90 (2014) 122005.
\newblock \href {http://arxiv.org/abs/1409.4809} {\path{arXiv:1409.4809}}.

\bibitem{PAO2015a}
A.~Aab, et~al., {The Pierre Auger Cosmic Ray Observatory}, Nucl. Instrum. Meth.
  A798 (2015) 172--213.
\newblock \href {http://arxiv.org/abs/1502.01323} {\path{arXiv:1502.01323}},
  \href {http://dx.doi.org/10.1016/j.nima.2015.06.058}
  {\path{doi:10.1016/j.nima.2015.06.058}}.

\bibitem{Gilmore2012}
R.~C. {Gilmore}, R.~S. {Somerville}, J.~R. {Primack}, A.~{Dom{\'{\i}}nguez},
  {Semi-analytic modelling of the extragalactic background light and
  consequences for extragalactic gamma-ray spectra}, Monthly Notices of the
  Royal Astronomical Society 422 (2012) 3189--3207.
\newblock \href {http://arxiv.org/abs/1104.0671} {\path{arXiv:1104.0671}},
  \href {http://dx.doi.org/10.1111/j.1365-2966.2012.20841.x}
  {\path{doi:10.1111/j.1365-2966.2012.20841.x}}.

\bibitem{Koning2005}
A.~J. Koning, S.~Hilaire, M.~C. Duijvestijn, Talys: Comprehensive nuclear
  reaction modeling, AIP Conference Proceedings 769~(1) (2005) 1154--1159.
\newblock \href {http://dx.doi.org/http://dx.doi.org/10.1063/1.1945212}
  {\path{doi:http://dx.doi.org/10.1063/1.1945212}}.

\bibitem{AlvesBatista2015a}
R.~Alves~Batista, D.~Boncioli, A.~di~Matteo, A.~van Vliet, D.~Walz, {Effects of
  uncertainties in simulations of extragalactic UHECR propagation, using
  CRPropa and SimProp}, JCAP 1510~(10) (2015) 063.
\newblock \href {http://arxiv.org/abs/1508.01824} {\path{arXiv:1508.01824}},
  \href {http://dx.doi.org/10.1088/1475-7516/2015/10/063}
  {\path{doi:10.1088/1475-7516/2015/10/063}}.

\bibitem{Walz2016}
D.~Walz, Constraining models of the extragalactic cosmic-ray origin with the
  pierre auger observatory, Ph.D. thesis, RWTH Aachen University (2016).

\bibitem{Werner2011}
K.~Werner, I.~Karpenko, T.~Pierog, ``ridge'' in proton-proton scattering at 7
  tev, Phys. Rev. Lett. 106 (2011) 122004.
\newblock \href {http://dx.doi.org/10.1103/PhysRevLett.106.122004}
  {\path{doi:10.1103/PhysRevLett.106.122004}}.

\bibitem{DeDomenico2013a}
M.~De~Domenico, M.~Settimo, S.~Riggi, E.~Bertin, {Reinterpreting the
  development of extensive air showers initiated by nuclei and photons}, JCAP
  1307 (2013) 050.
\newblock \href {http://arxiv.org/abs/1305.2331} {\path{arXiv:1305.2331}},
  \href {http://dx.doi.org/10.1088/1475-7516/2013/07/050}
  {\path{doi:10.1088/1475-7516/2013/07/050}}.

\bibitem{Patil2010}
A.~Patil, D.~Huard, C.~Fonnesbeck, Pymc: Bayesian stochastic modelling in
  python, Journal of Statistical Software 35~(1) (2010) 1--81.
\newblock \href {http://dx.doi.org/10.18637/jss.v035.i04}
  {\path{doi:10.18637/jss.v035.i04}}.

\bibitem{Gelman1992}
A.~Gelman, D.~B. Rubin, Inference from iterative simulation using multiple
  sequences, Statistical Science 7~(4) (1992) pp. 457--472.

\bibitem{Aloisio2011}
R.~Aloisio, V.~Berezinsky, A.~Gazizovb, Ultra high energy cosmic rays: The
  disappointing model., Astroparticle Physics 34 (2011) 620--626.
\newblock \href {http://arxiv.org/abs/0907.5194} {\path{arXiv:0907.5194}},
  \href {http://dx.doi.org/10.1016/j.astropartphys.2010.12.008}
  {\path{doi:10.1016/j.astropartphys.2010.12.008}}.

\bibitem{Boncioli2015}
D.~Boncioli, A.~di~Matteo, A.~Grillo, {Surprises from extragalactic propagation
  of UHECRs}, in: {Cosmic Ray International Seminar: The status and the future
  of the UHE Cosmic Ray Physics in the post LHC era (CRIS 2015) Gallipoli,
  Italy, September 14-16, 2015}, 2015.
\newblock \href {http://arxiv.org/abs/1512.02314} {\path{arXiv:1512.02314}}.

\bibitem{Porcelli2015}
A.~Porcelli, {for the Pierre Auger Collaboration}, Measurements of $x_{\max}$
  above $10^{17}$ ev with the fluorescence detector of the pierre auger
  observatory, in: Proceedings of the 34th International Cosmic Ray Conference,
  2015.

\bibitem{Buitink2016}
S.~Buitink, et~al., A large light-mass component of cosmic rays at 1017-1017.5
  electronvolts from radio observations, Nature 531~(7592) (2016) 70--73.

\bibitem{Thoudam2016}
S.~Thoudam, J.~P. Rachen, A.~van Vliet, A.~Achterberg, S.~Buitink, H.~Falcke,
  J.~R. Hörandel, {Cosmic-ray energy spectrum and composition up to the ankle
  - the case for a second Galactic component}, Astronomy \& Astrophysics under
  review.
\newblock \href {http://arxiv.org/abs/1605.03111} {\path{arXiv:1605.03111}}.

\bibitem{Hillas1984}
A.~M. Hillas, The origin of ultra-high-energy cosmic rays, Annual Review of
  Astronomy and Astrophysics 22 (1984) 425--44.
\newblock \href {http://dx.doi.org/10.1146/annurev.aa.22.090184.002233}
  {\path{doi:10.1146/annurev.aa.22.090184.002233}}.

\bibitem{Ptitsyna2010}
K.~V. Ptitsyna, S.~V. Troitsky, Physical conditions in potential accelerators
  of ultra-high-energy cosmic rays: updated hillas plot and radiation-loss
  constraints, Physics-Uspekhi 53~(7) (2010) 691.
\newblock \href {http://arxiv.org/abs/0808.0367} {\path{arXiv:0808.0367}},
  \href {http://dx.doi.org/10.3367/UFNe.0180.201007c.0723}
  {\path{doi:10.3367/UFNe.0180.201007c.0723}}.

\bibitem{Peterson2006}
J.~R. Peterson, A.~C. Fabian, {X-ray spectroscopy of cooling clusters}, Phys.
  Rept. 427 (2006) 1--39.
\newblock \href {http://arxiv.org/abs/astro-ph/0512549}
  {\path{arXiv:astro-ph/0512549}}, \href
  {http://dx.doi.org/10.1016/j.physrep.2005.12.007}
  {\path{doi:10.1016/j.physrep.2005.12.007}}.

\bibitem{Lemoine2009}
M.~{Lemoine}, E.~{Waxman}, {Anisotropy vs chemical composition at ultra-high
  energies}, JCAP 11 (2009) 009.
\newblock \href {http://arxiv.org/abs/0907.1354} {\path{arXiv:0907.1354}},
  \href {http://dx.doi.org/10.1088/1475-7516/2009/11/009}
  {\path{doi:10.1088/1475-7516/2009/11/009}}.

\bibitem{Abreu2013a}
P.~Abreu, et~al., {Bounds on the density of sources of ultra-high energy cosmic
  rays from the Pierre Auger Observatory}, JCAP 1305 (2013) 009.
\newblock \href {http://arxiv.org/abs/1305.1576} {\path{arXiv:1305.1576}},
  \href {http://dx.doi.org/10.1088/1475-7516/2013/05/009}
  {\path{doi:10.1088/1475-7516/2013/05/009}}.

\bibitem{Aab2015f}
A.~Aab, et~al., Search for patterns by combining cosmic-ray energy and arrival
  directions at the pierre auger observatory 75~(6) (2015) 269.
\newblock \href {http://arxiv.org/abs/1410.0515} {\path{arXiv:1410.0515}},
  \href {http://dx.doi.org/10.1140/epjc/s10052-015-3471-0}
  {\path{doi:10.1140/epjc/s10052-015-3471-0}}.

\bibitem{Aartsen2016}
M.~G. Aartsen, et~al., {All-sky search for time-integrated neutrino emission
  from astrophysical sources with 7 years of IceCube data}, The Astrophysical
  Journal under review.
\newblock \href {http://arxiv.org/abs/1609.04981} {\path{arXiv:1609.04981}}.

\bibitem{Anchordoqui1999}
L.~A. Anchordoqui, G.~E. Romero, J.~A. Combi, {Heavy nuclei at the end of the
  cosmic ray spectrum?}, Physical Review D 60 (1999) 103001.
\newblock \href {http://arxiv.org/abs/astro-ph/9903145}
  {\path{arXiv:astro-ph/9903145}}, \href
  {http://dx.doi.org/10.1103/PhysRevD.60.103001}
  {\path{doi:10.1103/PhysRevD.60.103001}}.

\end{thebibliography}

\end{document}